\documentclass[fleqn,usenatbib]{mnras}

\usepackage{newtxtext,newtxmath}
\usepackage{xspace}

\usepackage{natbib}  
\setcitestyle{round,aysep={},yysep={;}}  

\usepackage{graphicx}	
\usepackage{amsmath}	
\usepackage{float}

 \title{Hubble tensions: a historical statistical analysis}
\author[L\'opez-Corredoira]{Mart\'\i n L\'opez-Corredoira$^{1,2}$\thanks{E-mail: martin@lopez-corredoira.com}
\\
$^1$ Instituto de Astrof\'\i sica de Canarias, E-38205 La Laguna, Tenerife, Spain\\
$^2$ Departamento de Astrof\'\i sica, Universidad de La Laguna,
E-38206 La Laguna, Tenerife, Spain
}

\date{Last Rev. 29 July 2022}

\begin{document}
\label{firstpage}
\maketitle

\begin{abstract}
Statistical analyses of the measurements of the Hubble-Lema\^itre constant $H_0$ (163 measurements between 1976 and 2019) show that
the statistical error bars associated with the observed parameter measurements have been underestimated---or the systematic errors were not properly taken into account---in at least 15-20\% of the measurements. 
The fact that the underestimation of error bars for $H_0$ is so common might explain the apparent discrepancy of values, which is 
formally known today as the Hubble tension.
  Here we have carried out a recalibration of the probabilities with this sample of measurements.
 We find that $x\sigma $ deviation is indeed equivalent in a normal 
distribution to $x_{\rm eq.}\sigma $s deviation in the frequency of values, 
where $x_{\rm eq.}=0.83x^{0.62}$.
Hence, a tension of 4.4$\sigma $, estimated between the local Cepheid--supernova distance ladder and cosmic microwave background (CMB) data, is indeed a 2.1$\sigma $ tension in equivalent terms of a normal distribution of frequencies, with an associated probability $P(>x_{\rm eq.})=0.036$ (1 in 28). This can be increased up to a equivalent tension of 2.5$\sigma $ in the worst of the cases of claimed 6$\sigma $ tension, 
which may anyway happen as a random statistical fluctuation.
\end{abstract}


   \maketitle
%

\section{Introduction}
\label{.intro}

The Hubble--Lema\^itre constant, $H_0$, is one of the fundamental cosmological parameters.
We know its value approximately, but there is not yet a consensus regarding an accurate estimation of the parameter.

From the beginning of the 
discovery of the apparent magnitude relation of the galaxies in the 1920s, first by Lema\^itre
and later by Hubble (the so-called Hubble--Lema\^itre diagram), $H_0$ has decreased 
the value of that constant by almost an order of magnitude. 
In the 1980s, two preferred values were defended by 
different teams: either 50 or 100 km s$^{-1}$ Mpc$^{-1}$. Later, in the 1990s and 2000s, a value of around 70 km s$^{-1}$ Mpc$^{-1}$ became dominant, with preference for the value of 72 
km s$^{-1}$ Mpc$^{-1}$ obtained by the Hubble Space Telescope (HST) Key Project using supernovae  \citep{Fre10}. 
Nonetheless, discordant values were later published. A period--luminosity 
bias for the calibration of distances with nearby galaxies would justify a
Hubble--Lema\^itre constant to be reduced to values of around 60 km s$^{-1}$ Mpc$^{-1}$ \citep{Pat08,San09}. Even supernova data with HST were fitted with these values. 
As to the possible (non-)universality of the Cepheid period--luminosity relation, 
it was argued that low metallicity Cepheids have flatter slopes, 
and that the derived distance would depend on what relation is used \citep{Tam03}. 

We must also bear in mind that the value of $H_0$ is determined without knowing on which scales the radial motion of galaxies and clusters of galaxies relative to us is completely 
dominated by the Hubble--Lema\^itre flow \citep{Mat95}. The homogeneity scale may be much larger than expected \citep{Yad10,Syl11}, thus giving important net velocity flows on large scales that are incorrectly attributed to cosmological 
redshifts. Also, values of $H_0$ derived from cosmic microwave background radiation (CMBR) analyses are subject to the errors in the cosmological interpretation of this radiation \citep{Lop13}. That is, CMBR data are interpreted under the assumption of some model, which relates $\Lambda $CDM cosmological model with the power spectrum of CMBR, and there is the possibllity that CMBR can be interpreted with a different model. For instance, a Friedmann-Lema\^itre-Robertson-Walker metric lacking dark energy would also produce a big change in the values of $H_0$ \citep{Bla03}. Moreover, Galactic foregrounds are not perfectly removed \citep{Lop07,Axe15,Cre21}, and these are another source of uncertainties.
 
Yet another controversy arose more recently on the value of $H_0$. The Hubble--Lema\^itre constant estimated from the local Cepheid--supernova distance ladder is
at odds with the value extrapolated from CMB data, assuming the standard cosmological model,
$74.0\pm 1.4$ \citep{riess2019large} and $67.4\pm 0.5$ km s$^{-1}$ Mpc$^{-1}$ \citep{Pla18} respectively, which gives an incompatibility at the 4.4$\sigma $ level. This tension can even be increased up to 6$\sigma $ depending on the datasets considered \citep{DiV21}. 
Given the number of systematic errors that may arise in the measurements, this should not be surprising. However, this problem has motivated hundreds of papers since 2019 and
many solutions to the problem have been proposed---see the reviews in \citet{DiV21,Per21,Abd22})---discussing either the method to estimate $H_0$ or new theoretical scenarios. 

Here we will not contribute with a new solution to this Hubble tension in physical terms. 
Instead we carry out
an historical investigation to  determine whether or not the given error bars truly represented the dispersion of values in a historical compilation of $H_0$ values. We also show 
how we can use this knowledge to recalibrate the probabilities of some tension of this kind.

\section{Bibliographical data and statistical analysis}

We use the compililation of 163 values for $H_0$ between the years of 1976 and 2019 by \citet{Fae20}, where 120 measurements between 1990-2010 were taken from the previous compilation 
by \citet{croft2015measurement},\footnote{All of the data
except the measurement of $93\pm 1$ km s$^{-1}$ Mpc$^{-1}$ by \citet{Chi95}, which we excluded for being
$>20\sigma $ away from the average.} plus other 8 measurements between 1976 and 1989 and 35 measurements between 2011 and 2019.
\citet{croft2015measurement} have made use of the NASA Astrophysics Data System
to generate the dataset by carrying out an automated search
of publication abstracts for the years 1990-2010, limiting the search to published papers which include 
values of $H_0$ and their error bars in the paper abstract
itself. According to the authors, although this selection does not cover the 100\% of the whole literature, it 
gathers most the measurements; this is representative and should not produce any statistical bias. 
The extra 43 measurements in \citet{Fae20} obey a similar selection method.

The latest measurement we use is the value given by \citet{riess2019large}, which marked the origin of the present-day Hubble tension. Our goal is investigating the historical records previous to 2019, and not the most recent ones,
because we want to investigate how common the kind of tension pointed out by \citet{riess2019large} was until then.
We wonder whether a 4.4$\sigma $ is something never previously seen or something ordinary. The research of data after \citet{riess2019large}'s paper in 2019 is not the scope of this paper, as this would belong to an analysis of the post-Hubble
tension epoch.

The~correlation factor of $H_0$ with time\footnote{For two independent variables $X$ and $Y$,  the~correlation factor 
is defined as $c=\frac{\langle X\, Y\rangle}{\langle X\rangle \langle Y\rangle}-1$, with~error $Err(c)=\frac{\sigma _X\sigma _Y}{\sqrt{N}\langle X\rangle \langle Y\rangle}$. 
The Pearson correlation coefficient would be $\frac{c}{\sqrt{N}Err(c)}$.}  
is $c=0.027\pm 0.013$, a~$2\sigma $ significant correlation \citep{Fae20}.
Two sigma correlation is not a high significant detection, it is possibly a statistical fluctuation.
If a more significant correlation had been obtained, it would have been a proof that the measurements of
the parameter are not independent and there are systematic common errors variable with time or confirmation biases.\footnote{In his famous lecture on `Cargo Cult Science' (1974), Richard Feynman gave an example of Nobel Laureate Robert Millikan measuring the charge of an electron: ``it's interesting to look at the history of measurements of the charge of the electron, after Millikan. If you plot them as a function of time, you find that one is a little bigger than Millikan's, and the next one is a little bit bigger than that, and the next ones a little bit bigger than that, until finally they settle down to a number which is higher''\citep{Fey74}. Feynman goes on to ask why the final higher number was not discovered right away, and comes to the conclusion that when ``[scientists] got a number that was too high above Millikan's, they thought something must be wrong
and they would look for and find a reason why something might be wrong, leading them to eliminate values that were too far off, and did other things like that''.}

\subsection{Analysis of 1976-2019 data}

We neglect the variation with time of $H_0$, and we continue our analysis with the
weighted average.
The~weighted averages of the parameters in question were calculated by weighting each point by the inverse 
variance of that value:
\begin{equation}
\chi ^2=\sum _{i=1}^N\frac{(H_{0,i}-\overline{H_0})^2}{\sigma _i^2} 
.\end{equation} 
We obtain $\overline{H_0}=68.26\pm 0.40$ km s$^{-1}$ Mpc$^{-1}$ and  $\chi^2=575.7$ \citep{Fae20}.
For the value of $\chi^2$ calculated using the weighted average, 
the probability that the observed trend is due to chance is $Q = 1.0 \times 10^{-47}$. 
In order to reach a value for $Q$ that is statistically significant ($Q \geq 0.05$), 27 outliers (with more than 2.8$\sigma $ deviation from the average value) must be removed from the data ($n = 136$, $\chi^2=161.3$), producing a value for $Q$ of 0.061. 
These numbers are slightly different from those given by \citet{Fae20}, due to a minor error
correction. Table \ref{Tab:badvalues} lists the 27 values with more than 2.8$\sigma $ deviation.
If instead of 68.26 km s$^{-1}$ Mpc$^{-1}$ we took another value, the $\chi^2$ would be larger, and
the number of points to reject to get $Q \geq 0.05$ would also be larger.

Because of the simplifying assumption made about the neglect of the covariance of each observed measurement, 
the ratio of 27 in 163 is an approximation. A considerable fraction of the published measurements were not independent at all, but consisted of incremental updates to a chain of studies often based on a common foundation of measurements and assumptions.
In general, treating the data as independent variables, as we do here, gives a conservative limit of the
real dispersion of data, because the 'effective' number of degrees of freedom is lower than the number assuming independence, thus increasing the reduced $\chi^2 $ (i.e., decreasing even more the probability
of this distribution of measurements). Nonetheless, there may be exceptions to this rule: for instance, if
we take only two measurements with a large tension between them, and we multiply each of the points
$n$ times, the obtained $Q$ with $2n$ points would be lower. However, this would happen only if the number
of points that are multiplied are precisely those with a priori tension; if we multiple the points selected randomly, the
value of $Q$ obtained would be larger than the original value. Therefore, a very low $Q$ might
increase its value, but never within a range of $Q\geq 0.05$. That is, if we find a general tension in the data
indicated by very low values of the probability $Q$, this tension would not be erased by considerations of covariance
term analyses.

\begin{table}
\caption{Outliers: measurements of $H_0$ in which $|H_0-\overline{H_0}|>2.8\sigma $, where $\overline{H_0}=68.26$ km s$^{-1}$ Mpc$^{-1}$ is the weighted average of the 163 values of the literature.}
\label{Tab:badvalues}
\begin{center}
\begin{tabular}{cccc}
Year & $H_0$ (km s$^{-1}$ Mpc$^{-1}$) & $\frac{|H_0-\overline{H_0}|}{\sigma }$ & Authors  \\ \hline
   1976 &      $50.3\pm 4.3$  &      4.2       &  Sandage \& Tammann \\
   1984 &       $45.0\pm 7.0$ &      3.3       &  J{\~o}eveer  \\
   1990 &       $52.0 \pm 2.0$ &     8.1       &   Sandage \& Tammann \\ 
   1993 &       $47.0 \pm 5.0$ &     4.3       &   Sandage \& Tammann \\ 
   1994 &       $85.0 \pm 5.0$ &     3.3       &   Lu et al. \\ 
   1996 &       $84.0 \pm 4.0$ &     3.9         &  Ford et al. \\        
   1996 &       $57.0 \pm 4.0$ &     2.8        &  Branch et al. \\   
   1996 &       $56.0 \pm 4.0$ &     3.1       &    Sandage et al.\\   
   1998 &       $65.0 \pm 1.0$ &     3.3       &  Watanahe et al. \\  
   1998 &       $44.0 \pm 4.0$ &     6.1     &  Impey et al. \\  
   1999 &       $60.0 \pm  2.0$ &    4.1        &  Saha et al. \\  
   1999 &       $55.0 \pm  3.0$ &    4.4   &  Sandage \\   
   1999 &       $54.0 \pm  5.0$ &    2.9      &  Bridle et al. \\
   1999 &       $42.0 \pm 9.0$ &     2.9    &  Collier et al. \\  
   2000 &       $65.0 \pm 1.0$ &     3.3     &  Wang et al. \\
   2000  &      $52.0\pm  5.5$ &     3.0     &   Burud et al. \\
   2004  &      $78.0\pm  3.0$ &     3.2     &   Wucknitz et al. \\ 
%
   2006  &      $74.9\pm 2.3$  &     3.0      &  Ngeow \& Kanbur \\  
   2006  &      $74.0\pm 2.0$  &    2.9      &   S\'anchez et al. \\
   2008  &      $61.7\pm 1.2$  &  5.7        &  Leith et al. \\
   2012  &      $74.3\pm 2.1$  & 2.9 &  Freedman et al.\\
   2013  &      $76.0\pm 1.9$  &  4.1 &  Fiorentino et al. \\
   2016  &      $73.2\pm 1.7$  &  2.9 &  Riess et al. \\
   2018  &      $73.5\pm 1.7$  &  3.1 &  Riess et al. \\
   2018  &      $73.3\pm 1.7$  &  3.0 &  Follin \& Knox \\
   2018  &      $73.2\pm 1.7$  & 2.9  & Chen et al.  \\
   2019  &      $74.0\pm 1.4$  & 4.1 & Riess et al. \\ \hline
\end{tabular}
\end{center}
\end{table}

\subsection{Analysis of the 2001-2019 data}

The enormous discrepancies between cited uncertainties in derived values of $H_0$ and the much larger dispersions when values from different authors were compared to each other was already realized 
in the 1980s and 1990s \citep[e.g.,][]{Han82,Row85,Ken95}.
We now know that there were several underlying reasons for this underestimation of uncertainties in $H_0$ in the measurements before the year $\approx 2000$: errors were usually estimated rather than measured directly via comparison of independent methods; and systematic errors were usually ignored and even the statistical
errors were underestimated. 
Around the turn of the century, when beginning, for example, with the Hubble Space Telescope Key Project on the distance scale, authors began quantifying their uncertainties via multiple independent distance measurements, with separate explicit, and conservative, estimates of the remaining systematic errors 
\citep[e.g.,][]{Fre01}. 
However, looking at the values at Table \ref{Tab:badvalues}, we see many points later than the year 2000 in the list of 
outliers, which have $>2.8\sigma $ deviation. Certainly, the absolute values of $H_0$ measurement after the year 2000 are closer to the value of the weighted average, because the error bars are also smaller than those ones before 2000. However, the important thing here is not whether the error bars are lower, but whether the error bars are accurately measured, and we have a bunch of values measured after 2000 that are not within this category, assuming that $\overline{H_0}=68.26$ km s$^{-1}$ Mpc$^{-1}$ represents the real value.

We may think that the value of $\overline{H_0}=68.26$ km s$^{-1}$ Mpc$^{-1}$ is not correct, since it was
obtained with the weighted average of the whole sample 1976-2019. We repeat the whole analysis only
with the subsample of 2001-2019 data: they are 85 measurements.
For this, we obtain $\overline{H_0}=69.44\pm 0.26$ km s$^{-1}$ Mpc$^{-1}$ and  $\chi^2=195.2$.
The probability that the observed distribution is due to chance is $Q = 7.7 \times 10^{-11}$. 
In~order to reach a value for $Q$ that is statistically significant ($Q \geq 0.05$), at least
six outliers (with more than 2.8$\sigma $ deviation from the average value) must be removed from the data ($n = 79$, $\chi^2=94.8$), producing a value for $Q$ of 0.095. 
Table \ref{Tab:badvalues2} lists the six measurements with more than 2.8$\sigma $ deviation.
The conclusion is that the data of 2001-2019 are significantly better (only 7\% of data need to be 
rejected to get a probability $Q>0.05$), but there are also tensions here. This includes the recent
tension of CMBR vs. supernovae data given from \citet{riess2019large} within $\Lambda $CDM cosmology, 
but there are also other measurements with unacceptable high
devitations from the average. The fact there is less relative deviation in more recent data might
be due to the non-independence of the data; if we took truly independent measurements, using different
techniques, this ratio would presumably be much higher.

We might think that the problem of a too low $Q$ in this last analysis with the subsample 2001-2019
can be avoided if we remove some points we suspect to be be wrong, for instance the value
of $61.7\pm 1.2$ km s$^{-1}$ Mpc$^{-1}$ given by \citet{Lei08} using SNe Ia. However, removing some values because we do not like them and do not agree the dominant trends is not a valid objective method of data analysis. We can see that there were important Hubble tensions long before the announcement of Riess et al.
and a rigurous statistical analysis cannot decide that some values are more trustable than others.
\citet{Lei08} might have underestimated their errors, and \citet{riess2019large} might also have
underestimated the errors. Saying that we trust one team more than another is not a scientific approach.

As said in the previous subsection, changing the weighted average ($69.44\pm 0.26$ km s$^{-1}$ Mpc$^{-1}$ herein) for another value does not solve anything, but makes the statistics worse.
For instance, if we choose as reference the value of 74.0 km s$^{-1}$ Mpc$^{-1}$ given by 
\citet{riess2019large}, we obtain a $\chi^2=561.9$ for the 85 points (85 degrees of freedom), with
associated $Q = 2.4 \times 10^{-71}$, and
in~order to reach a $Q \geq 0.05$, we would have to remove at least eight measurements.

\begin{table}
\caption{Outliers: measurements of $H_0$ in which $|H_0-\overline{H_0}|>2.8\sigma $, where $\overline{H_0}=69.44$ km s$^{-1}$ Mpc$^{-1}$ is the weighted average of the 85 values of the literature between (included) 2001 and 2019.}
\label{Tab:badvalues2}
\begin{center}
\begin{tabular}{cccc}
Year & $H_0$ (km s$^{-1}$ Mpc$^{-1}$) & $\frac{|H_0-\overline{H_0}|}{\sigma }$ & Authors  \\ \hline
2002 & $44.0\pm 9.0$ & 2.8 &  Winn et al. \\ 
2004  &      $78.0\pm  3.0$ &     2.9     &   Wucknitz et al. \\ 
2008  &      $61.7\pm 1.2$  &  6.7        &  Leith et al. \\
   2013  &      $76.0\pm 1.9$  &  3.5 &  Fiorentino et al. \\
   2018  &      $67.4\pm 0.5$ &   4.1 &   Chen et al.           \\ 
   2019  &      $74.0\pm 1.4$  & 3.2  & Riess et al. \\ \hline
\end{tabular}
\end{center}
\end{table}

\section{Recalibration of probabilities}

In Fig. 1, we plot the frequency of deviation larger than $x\sigma$ 
from the weighted average value $H_{0,{\rm teor.}}=68.26$ km s$^{-1}$ Mpc$^{-1}$ derived
using the whole sample of 163 measurements (including the outliers).
Clearly, the probabilities are much higher than those expected in a normal Gaussian error distribution.
For example, in a Gaussian error distribution we should get a $P=2.7\times 10^{-3}$ of obtaining
a deviation higher than 3$\sigma $ (where $\sigma $ is the error of the measurement), but instead we observe that 11.7\% of our measurements get deviations
higher than 3$\sigma $.
The fit of our probability with the frequencies we obtain from the real measurements is:
\begin{equation}
\label{probeq}
P(|H_0-\overline {H_0}| >x\, \sigma )=(0.93\pm 0.06)\times exp[-(0.720\pm 0.013)x]
.\end{equation}
This is equivalent (fit in the range of $x$ between 1 and 12) 
to a number of $\sigma $s deviation in a normal
distribution
(see Fig. \ref{Fig:equiv}):
\begin{equation}
x_{\rm eq.}=(0.830\pm 0.004)x^{0.621\pm 0.003}
,\end{equation}
where $x$ is the number of $\sigma $s in the measurement (i.e. $x=\frac{|H_0-\overline{H_0}|}{\sigma }$).
Hence, for instance, a datum that is 3.0$\sigma $ away for the expected value should not be
interpreted as a 3.0$\sigma $ tension, but a 1.6$\sigma $ one in equivalent terms of a normal distribution ($P(>x_{\rm eq.})=0.11$). Likewise, a tension of 4.4$\sigma $ (as for instance claimed by \citet{riess2019large}) for a Hubble tension) is indeed a 2.1$\sigma $ tension in equivalent terms of a normal distribution, with an associated $P(>x_{\rm eq.})=0.036$ (1 in 28), which is not large but it can occur as a random statistical fluctuation.
For an even larger limit of the tension, at $6\sigma $, as pointed by 
using some different datasets \citep{DiV21}, we would have a equivalent 2.5$\sigma $
with an associate $P(>x_{\rm eq.})=0.012$ (1 in 83), which is still not amazing.

\begin{figure}
\centering\includegraphics[width=1.0\linewidth]{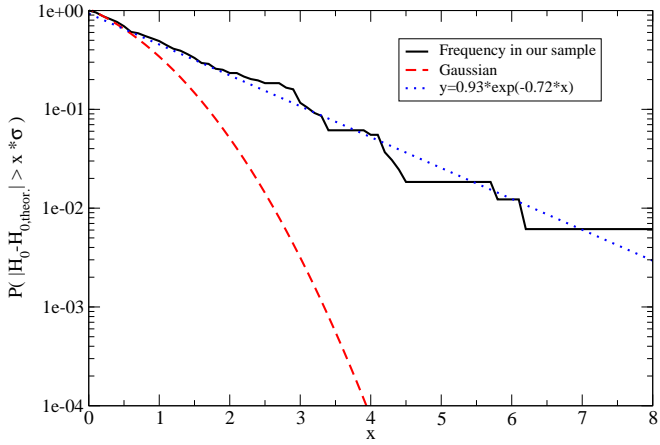}
\caption{Probability of deviation larger than $x$ sigmas, 
assuming $H_{0,{\rm teor.}}=68.26$ km s$^{-1}$ Mpc$^{-1}$ (the weighted average value). The dotted line shows the best exponential fit. The dashed line denotes
the expected probability if the errors were Gaussian.}
\label{Fig:prob}
\end{figure}

\begin{figure}
\centering\includegraphics[width=1.0\linewidth]{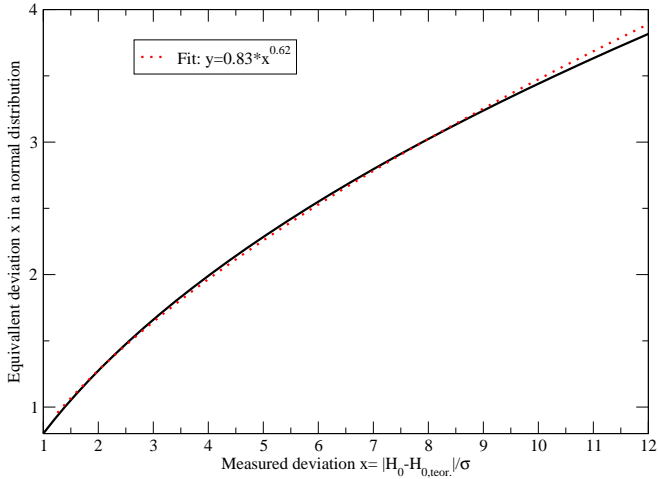}
\caption{Equivalent number of $\sigma $ of a normal distribution as a function of 
the number of measured $\sigma $ of deviation with respect the weighted
average value of $H_0$, assuming that the probababilities are given by the
blue dotted line of Fig. \ref{Fig:prob} (Eq. \ref{probeq}).}
\label{Fig:equiv}
\end{figure}

\section{Conclusions}

We have examined the trend and dispersion of values of the meaurements of $H_0$ during 43 years (1976-2019). 
The probabilities $Q$ for the distribution of different measurements of $H_0$ and their errors
are extremely low, both with respect to a constant value (weighted average of all the measurements)
or with a linear fit \citep{Fae20}. We need removing 24-27 measurements 
to reach a statistically significant dataset ($Q\geq 0.05$).

In~the light of the analysis carried out here, one would not be surprised to find cases like
the 4.4$\sigma$ discrepancy seen between the best measurement using Supernovae Ia in \citet{riess2019large}. It is likely that the underestimation of error bars for $H_0$ in many measurements contributes to the apparent 4.4$\sigma$ discrepancy.
Here we have carried out a recalibration of the probabilities with the present sample of measurements
and we find that $x\sigma $ deviation is indeed equivalent in a normal 
distribution to the $x_{\rm eq.}\sigma $ deviation, where $x_{\rm eq.}=0.83x^{0.62}$.
Hence, the tension of 4.4$\sigma $, estimated between the local Cepheid--supernova distance ladder and cosmic microwave background (CMB) data, is indeed a 2.1$\sigma $ tension in equivalent terms of a normal distribution, with an associated $P(>x_{\rm eq.})=0.036$ (1 in 28). This is not 
large but it can occur as a random statistical fluctuation. This can be increased up to a equivalent tension of 2.5$\sigma $ in the worst of the cases.

One may be tempted to claim that the statistics and treatment of uncertainties in published $H_0$ measurements from the last century cannot be applied to the interpretation of the most recent measurements,
because we know the reason why the error bars measured $>20$ years ago were underestimated, whereas we
are totally certain that our present-day measurements of the error bar are accurate.
However, this statement would be a misunderstanding of the present analysis: we are not here discussing the physics and assumptions (as discussed in the introduction) 
behind those measurements, and certainly we do not have any argument to criticize the techniques of recent $H_0$ measurements. Rather, we are doing here a metaanalysis from a
historical point of view. Of course, Riess et al. think that their error bars are accurate, but can we be sure that within 30-40 years the same security can be maintained? We can also assume that Sandage, Tammann 
and other very high-qualified specialists measuring $H_0$ in the 1970s were also  convinced that their
measurements were correctly carried out, but it was later proven that they have underestimated their errors.
The idea that scientists in the past were not accurate enough in their statistical analyses
and present-day researchers are doing  a perfect job is not justified from an objective historical point of view.
In this sense, this investigation makes sense. The question is: how frequent was to get an underestimation of the error bars in $H_0$ between 1976 and 2019? And we got an answer, which is useful to understand present and future data analyses. 

\section*{Acknowledgements}
The author thanks Rupert Croft for providing data from his paper \citet{croft2015measurement}.
The author is also grateful to the anonymous referee for helpful comments.
\\

DATA AVAILABILITY: The data underlying this article are available in the article and the
cited references, and in its online supplementary material.
\\


\begin{thebibliography}{99}


\bibitem[Abdalla
  et~al.(2022)]{Abd22}
Abdalla E.,  Abell\'an G.~F.,  Aboubrahim A.,   et~al., 2022, arXiv.org, p.
  2203.06142

\bibitem[Axelsson et al.(2015)]{Axe15}
Axelsson M.,  Ihle H.~T.,  Scodeller S.,   Hansen F.~K.,  2015, Astron.
  Astrophys., 578, id. A44, 11 pp.

\bibitem[Blanchard et al.(2003)]{Bla03}
Blanchard A.,  Douspis M.,  Rowan-Robinson M.,   Sarkar S.,  2003, Astron.
  Astrophys., 412, 35

\bibitem[Chiba \& Yoshii(1995)]{Chi95}
Chiba M.,  Yoshii Y.,  1995, Astrophys. J., 442, 82

\bibitem[Creswell \& Naselsky(2021)]{Cre21}
Creswell J.,  Naselsky P.,  2021, J. Cosmol. Astropart. Phys., 2021(3), id. 103

\bibitem[Croft \& Dailey(2015)]{croft2015measurement}
Croft R. A.~C.,  Dailey M.,  2015, Quarterly Phys. Rev., 1, 1

\bibitem[Di Valentino, Mena  \& Pan(2021)]{DiV21}
Di Valentino E.,  Mena O.,   Pan S.,  2021, Class. Quantum Grav., 38, id.
  153001, 110 pp.


\bibitem[Faerber \& L\'opez-Corredoira(2020)]{Fae20}
Faerber T.,  L\'opez-Corredoira M.,  2020, Universe, 6, 114

\bibitem[Feynman(1974)]{Fey74}
Feynman R.~P.,  1974, Engineering and Science, 37, 10

\bibitem[Freedman \& Madore(2010)]{Fre10}
Freedman W.~L.,  Madore B.~F.,  2010, Annu. Rev. Astron. Astrophys., 48, 673

\bibitem[Freedman et~al.(2001)]{Fre01}
Freedman W.~L.,  Madore B.~F.,  Gibson B.~K.,   et~al., 2001, The Astrophysical
  Journal, 553, 47

\bibitem[Hanes(1982)]{Han82}
Hanes D.~A.,  1982, Monthly Notices of the Royal Astronomical Society, 201, 145

\bibitem[Kennicutt, Freedman  \& Mould(1995)]{Ken95}
Kennicutt R.~C. J.,  Freedman W.~L.,   Mould J.~R.,  1995, Astronomical
  Journal, 110, 1476

\bibitem[Leith, Ng  \& Wiltshire(2008)]{Lei08}
Leith B.~M.,  Ng S. C.~C.,   Wiltshire D.~L.,  2008, Astrophysical Journal
  Letters, 672, L91

\bibitem[L\'opez-Corredoira(2007)]{Lop07}
L\'opez-Corredoira M.,  2007, J. Astrophys. Astron., 28, 101

\bibitem[L\'opez-Corredoira(2013)]{Lop13}
L\'opez-Corredoira M.,  2013, Int. J. Mod. Phys. D, 22, id. 1350032, 18 pp.

\bibitem[Matravers, Ellis  \& Stoeger(1995)]{Mat95}
Matravers D.~R.,  Ellis G. F.~R.,   Stoeger W.~R.,  1995, Quaterly J. R.
  Astron. Soc., 36, 29

\bibitem[Paturel(2008)]{Pat08}
Paturel G.,  2008, in Baryshev Y.,  Taganov I.,   Teerikorpi P.,  eds, ,
  Practical Cosmology, vol. 2.
TIN, St.-Petersburg, pp 178--184

\bibitem[Perivolaropoulos \& Skara(2021)]{Per21}
Perivolaropoulos L.,  Skara F.,  2021, arXiv.org, p. 2105.05208

\bibitem[Planck Collaboration(2020)]{Pla18}
Planck Collaboration c., 2020, Astron. Astrophys., 641, A6

\bibitem[Riess et al.(2019)]{riess2019large}
{Riess} A.~G.,  {Casertano} S.,  {Yuan} W.,  {Macri} L.~M.,   {Scolnic} D.,
  2019, The Astrophysical Journal, 876, 85

\bibitem[Rowan-Robison(1985)]{Row85}
Rowan-Robison M.,  1985, Cosmological Distance Ladder: Distance and Time in the
  Universe.
W.H. Freeman \& Co Ltd., New York

\bibitem[Sandage, Tammann  \& Reindl(2009)]{San09}
Sandage A.,  Tammann G.~A.,   Reindl B.,  2009, Astron. Astrophys., 493, 471

\bibitem[Sylos~Labini(2011)]{Syl11}
Sylos~Labini F.,  2011, Class. Quantum Grav., 28, id. 164003

\bibitem[Tammann, Sandage  \& Reindl(2003)]{Tam03}
Tammann G.~A.,  Sandage A.,   Reindl B.,  2003, Astron. Astrophys., 404, 423

\bibitem[Yadav, Bagla  \& Khandai(2010)]{Yad10}
Yadav J.~K.,  Bagla J.~S.,   Khandai N.,  2010, Mon. Not. R. Astron. Soc., 405,
  2009

\end{thebibliography}

%

\end{document}